\documentclass{article}

\usepackage[english]{babel}

\usepackage[letterpaper,top=2cm,bottom=2cm,left=3cm,right=3cm,marginparwidth=1.75cm]{geometry}

\usepackage{amsmath}
\usepackage{graphicx}
\usepackage[colorlinks=true, allcolors=blue]{hyperref}
\usepackage{epigraph}

    \usepackage{etoolbox}
    \makeatletter
    \newlength\epitextskip
    \pretocmd{\@epitext}{\em}{}{}
    \apptocmd{\@epitext}{\em}{}{}
    \patchcmd{\epigraph}{\@epitext{#1}\\}{\@epitext{#1}\\[\epitextskip]}{}{}
    \makeatother

\title{Spacetime events from the inside out.}
\author{Gerard Milburn,\\
Mathematics and Physics, University of Sussex, UK.}

\begin{document}
\maketitle

\begin{abstract}
We argue that special and general theories of relativity implicitly assume spacetime events correspond to quantum  measurement outcomes. This leads to a change in how one should view the equivalence of spacetime and gravity.   We describe a Bell test using time-like measurements that indicates a non classical causal structure that does not violate no-signaling.  From this perspective, the violation of the Bell inequalities are already evidence for the non classical structure of flat spacetime as seen by an agent embedded in it. We argue that spacetime geometry can be learned by an embedded agent with internal actuators and sensors making internal measurements.   
\end{abstract}
\vspace*{\fill}
\epigraph{\textit{Physics alone may not be able to provide answers to the space and
time issue. Instead, it is up to neuroscience to address them.}\cite{Buzaski}} 

\section{Introduction.}

It has been apparent since at least the Chapel Hill conference in 1955\cite{Dewitt2011TheRO} that the world revealed by Schr\"{o}dinger and Heisenberg is incompatible with the world revealed by Einstein, despite the astounding experimental success of quantum theory and general relativity. The resulting discontent has fueled a search for a quantum theory of gravity despite the  lack of any compelling experimental evidence to do so. There is a general belief that a quantum understanding of space and time will emerge from a quantum theory of gravity. 

Or have we missed something? The experimental tests of quantum theory via Bell violations  are among the top-shelf achievements of quantum theory and quite consistent with the theory of relativity. Yet there is something mysterious about this consistency.  As Gisin puts it, "no story in space–time can describe nonlocal correlations"\cite{Gisin2013}. 

Gisin, like many others, describes the quantum correlations revealed in  Bell tests as  `nonlocal' correlations. It is hard to describe it any other way, yet it is easy to design expressions which violate the Bell inequalities even for time-like separated observers. In fact, the spacetime coordinates of the observers play no role at all. It is in this sense that that quantum correlations seem to be coming from outside spacetime itself.  What is at stake here is our deep-seated intuition that that an agent, like us, can only act {\em here and now} according to internal states. This perspective has been emphasised by Buzaski\cite{Buzaski}.

In this paper, we will discuss a version of the Bell tests that use a single agent and time-like separation of the detection events. One might think that this is unlikely to raise issues of non locality. However, using the same logic as the standard Bell violation argument, the single agent scenario suggests a surprising interpretation: retro causation without superluminal signalling to the past. 

Prior to quantum theory, we could describe the world as being built from classical variables that describe objective properties of that world. In quantum optics, for example, we perform experiments on the electromagnetic field using sources and detectors. The quantum field is not a classical field like Maxwell's fields. It does not have independent properties like electric or magnetic field magnitudes that have independent causal efficacy. Only measurements results in a particular experiment context have causal efficacy. The value of the electric field is established by a particular class of experiments (homodyne and heterodyne detection). If we want to estimate the intensity of the field, we do a very different kind of measurement; we count photons. These are complementary experiments in the sense of Bohr.  If we have a single charge in a superposition of two locations in an ion trap, we do not worry that the resulting EM field has no classical interpretation. We simply measure what QM predicts.

The classical gravitational field, as revealed in GR, is a very different matter. In Einstein's formulation gravity is represented by the same mathematical object as space-time geometry. To measure the gravitational field we need to find ways to estimate the metric. Einstein used imaginary clocks and rulers. Today, we use quantum field theory ... QED. The objective is to estimate the components of a metric as best we can given the bounds imposed by quantum uncertainty\cite{PhysRevD.96.105004}.  A good example is measurement of the perturbations to the Minkowski metric that define gravitational waves, $h_{\mu\nu}$. That is what LIGO does, and it certainly uses quantum fields. We tend to think that the gravitational field is an objective fact in the world as it is highly classical. We do not need to worry about graviton statistics. The analogue, in the context of the electromagnetic field, are those states excited by large classical current sources.

Recently,  experimentalists have started probing highly non classical sources of gravity. A simple example is a single massive object in a superposition of `two places' . This only makes sense if there is a background reference frame. The idea of `two places' presupposes a further background spacetime metric in addition to that produced by the gravitational field of the massive object itself. This is the analogue of the EM field of a charge in a superposition at two places in an ion trap.  We could simply claim that there is no objective field until we measure it. But there is a big difference. Gravity has the same mathematical structure as spacetime (the strong equivalence principle). Do we really want to say spacetime is not an objective feature of the world but only revealed, however obscurely, by particular measurement results?  This would require us to claim that spacetime events are comprehensively equivalent to measurement results and thus contingent on the kinds of measurement we make. What is at stake in these new experiments is not the reality of gravity or wether gravity is quantum or classical,  but the reality of spacetime. It must emerge as an observer-dependent feature by making measurements on something more fundamental. Is this an opening for a better spacetime story of the Bell violations?

\subsection{Bell tests with three agents.}
The standard description of photonic tests of Bell inequalities involves two agents and an entangled photon source. I will include an additional agent to make it clear how the required correlation functions are constructed. 

\begin{figure}[htbp]
\centering
\includegraphics[scale=0.5]{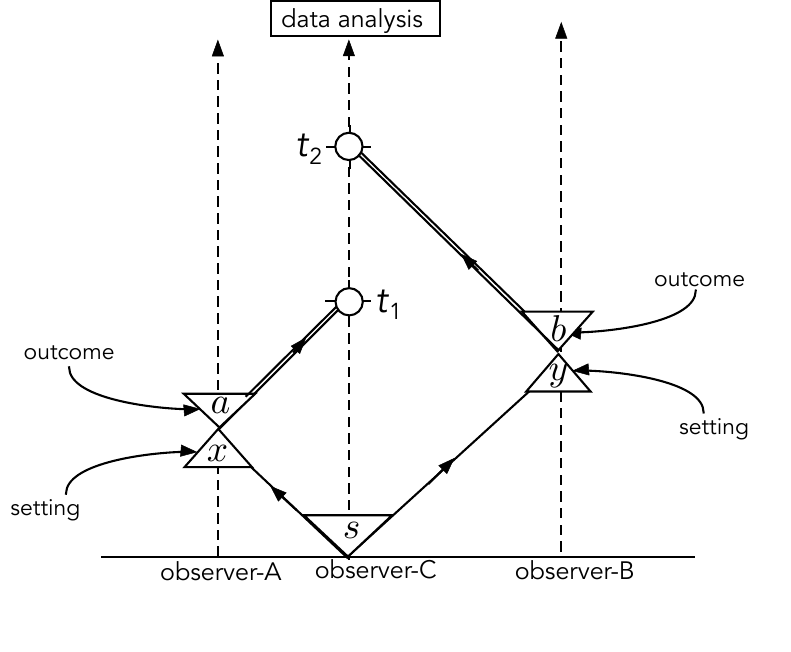}
\caption{A two-party Bell experiment with entangled photons.  A source at the origin produces pairs of entangled photons. The photons occupy oppositely directed spatial modes; one goes to detector-A and the other goes to detector-B. Both observers are space-lie separated. After each measurement, the setting and the outcome are sent, over classical channels, to a checker, observer-C, who stores the data for each trial and constructs the appropriate correlation function to check a Bell inequality.  }
\label{three-agent-Bell2}
\end{figure}

A standard Bell test with polarization-entangled photons is shown in Fig.(\ref{three-agent-Bell2}).  

In an ideal experiment, we use single-photon excitations of polarised spatio-temporal modes such that the photons are emitted in opposite directions, with anti-correlated polarisation states.   The photons are entangled in the polarisation. For example, the source could  prepare, in every trial,  the  two-photon state, 
\begin{equation}
	|\Psi^{-}\rangle = \frac{1}{\sqrt{2}}(|HV\rangle-|VH\rangle)
\end{equation}
where $|HV\rangle =|1\rangle_{k_A,H}\otimes|1\rangle_{k_B,V} $ and the subscripts label spatio-temporal modes with wave vectors $k_A,k_B$ and corresponding polarisation. A single trail corresponds to emitting a two-photon state and counting two photons;  one at observer-a and one at observer-B.  If for some reason two photons are not detected, that trial is discarded. As no detector is perfect this is likely to happen quite often, raising the detection loop-hole. 

Let the measurement settings be chosen from a set of discrete rotations $x \in \{\theta_{A,1}, \theta_{A,2} ,\ldots \theta_{A,n}\}$ and $y \in \{\theta_{B,1}, \theta_{B,2} ,\ldots \theta_{B,n}\}$. Suppose the detector settings are the same,  $x=y$, that is to say, the same angle is chosen.  We can rotate both angles jointly until we see a perfect anti-correlation at each output. When the photon at $A$ is detected at $a= 1$ channel the photon at $B$ is detected at $b= - 1$ channel, and vice versa.  However from trail to trial the local measurement outcomes are a random binary numbers $\pm 1$.

 In order to see the correlation, an observer must have access to both outcomes in each trial. In the lab this is obvious, as the experimentalist collects all the data from both detectors in each trial.  We make this explicit by introducing a `checker' --- labelled $C$ --- that receives the data (setting and outcome at each detector)  from each observer in a trial in the future light cone of the detection events. For convenience we will suppose the checker is at the same place as the source. Note that the checker receives purely classical information. A space-time diagram for the experiment is shown in Fig. (\ref{three-agent-Bell2}). Observer-C is at rest in the frame of the source and thus can easily synchronise emission and detection events to ensure that data is collected from the right photon pair in each trial.  

 The analysis of the data is well known. Typically, it involves computing a correlation function known as the CHSH correlation function. Classically this correlation function is bounded by $2$. Quantum  mechanics predicts that it is bounded by $2\sqrt{2}$ \cite{Scarani2019},  and many experiments have demonstrated that the classical bound is exceeded\cite{Gisin2013}.

\subsection{Bell tests with one agent.}
\begin{figure}[htbp]
\centering
\includegraphics[scale=0.5]{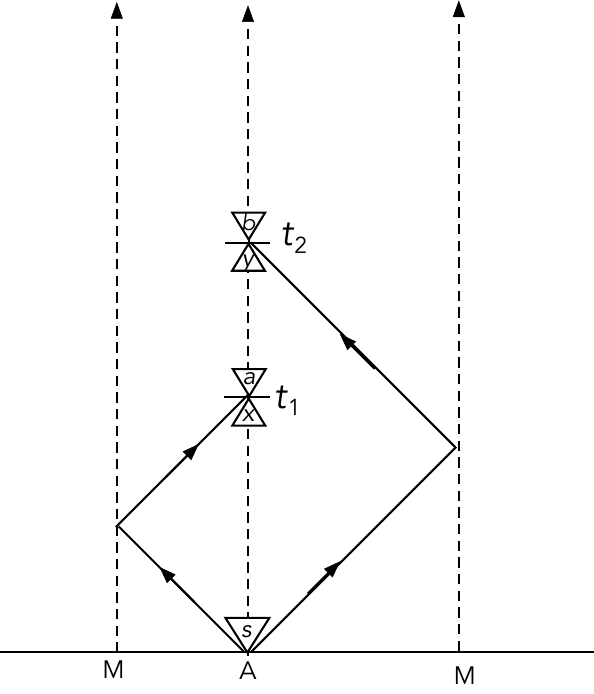}
\caption{A Bell test with a single observer/agent.   The agent  generates an entangled photon pair, in a known state (indicated by a joint measurement result $s$) and each photon travels in opposite directions where it is reflected by a mirror at rest in the agent's rest frame. One photon is received back by the agent at the early time $t_1$ and the other photon is received back at the later time $t_2$.  The dashed lines represent the world lines of the agent and the two mirrors. The variables $x,y$ represent the measurement settings and the variables $a,b$ represent the measurement outcomes as in a standard Bell test.  }
\label{one-observer}
\end{figure}

Consider the case depicted in Fig.(\ref{one-observer}). In this case we use the same source as in the usual Bell scenario but now mirrors, {\em asymmetrically} displaced either side of the agent,  reflect the photons back to a single agent at the source who will do the same single-photon experiments on each photon received in a given trial  to test a Bell violation. The single observer can easily store the classical results of the experiments and verify the violation of the CHSH inequalities at any point in the future of both measurements.

  Where is the puzzle of quantum entanglement for such an agent? Non locality is not an issue as all measurements are time-like separated.  Nevertheless, Bell inequalities will be violated if quantum theory is correct. As the quantum correlations are non signalling there can be no signalling from $t_1$ to $t_2$ or vice versa.   Yet there remains a deeper puzzle. If we tried to explain the correlations in terms of a local hidden variable, the logic of Bell's argument would imply a symmetric casual connection even without signalling. In other words, retro-causation without signalling.   We could explain the correlations as the measurement results at time $t_1$ causing the results at $t_2$ but we could equally claim that the measurement  at time $t_2$ caused the results at the earlier time, $t_1$. This is deeply at odds with our classical intuition.

\begin{figure}[h!]
	\centering
	\includegraphics[scale=0.5]{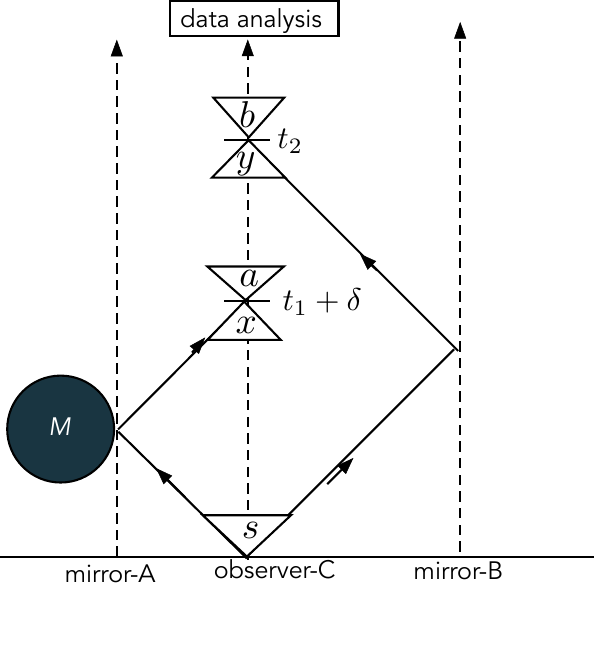}
 \caption{A single agent Bell test with gravity. }
 \label{gravity-one-observer}
	\end{figure}

The situation does not change in a gravitational field.  Consider the scheme shown in Fig. (\ref{gravity-one-observer}). This is a one agent protocol but one of the mirrors is replaced by a large mass. 
 The single photon pulse traveling to the left is blue-shifted going towards the mirror and red-shifted going away from the mirror. The net effect is simply a delay in the local detection time at the agent. This is the classical Shapiro shift\cite{shapiro}. It has no effect on the degree of violation of the Bell inequality. Even in curved spacetime the entanglement does not see classical gravity and the retro-causal interpretation remains. It is a purely local phenomenon.  Likewise, if the large mass is co-located with the agent at the origin.

\section{Quantum Field Theory and Spacetime Events. }
In Einstein's formulation, the gravitational field is determined by physical measurements made with clocks and rulers. Unfortunately Einstein was a little vague on just what he meant by local clocks and rulers, and even admitted that this was an inconsistency in the theory\cite{Einstein1970}.
\begin{quote}
	First, a critical remark on the theory as characterised above. It was noticeable that the theory introduces (besides four-dimensional space) two kinds of physical things, namely 1) rods and clocks, 2) all other things, e.g. the electromagnetic field, the material point, etc. This is in a sense inconsistent; rods and clocks should actually be presented as solutions to the basic equations (objects consisting of moving atomic structures), not as so to speak theoretically self-sufficient beings. However, the procedure is justified by the fact that it was clear from the beginning that the postulates of the theory are not strong enough to deduce from it sufficiently complete equations for physical events sufficiently free of arbitrariness to base a theory of rods and clocks on such a foundation. Unless one wanted to do without a physical interpretation of the coordinates altogether (which in itself would be possible), it was better to allow such inconsistencies---albeit with the obligation to eliminate them at a later stage of the theory. However, one must not legitimise the aforementioned sin to such an extent that one imagines that distances are physical beings of a special kind, essentially different from other physical quantities (``reducing physics to geometry,'' etc.).
\end{quote}
Einstein also made extensive use of classical light pulses to coordinate physical events.  We would now replace this with quantum fields. In \cite{PhysRevD.96.105004} various schemes were described for estimating spacetime metrics using quantum  fields and the ultimate accuracy achievable is determined by quantum uncertainty principles. Kempf has outlined a similar idea\cite{Kempf2021}. If we are deducing metrics from measurement outcomes then spacetime events are identified with measurement results taking place in some finite spacetime four volume. In a properly formulated quantum field theory,  all observers must agree on the probability distribution of such events.  Spacetime is objective if a little uncertain.   

In the case of flat spacetime, this approaches can easily identify an inertial reference frame, if we use semiclasssical states of light. However, what kind of spacetime are we to infer using the entangled states in the one-observer protocol described in the previous section? Identifying spacetime with these kinds of measurements is already implies a retro causal structure even in the case of no gravitational field, as we described in the previous section.  This would impact the causal set approach to quantum gravity\cite{causalset}. This is based on taking the causal structure of general relativity as axiomatic, although what is really meant by casual is in fact signalling.   The single observer Bell experiment suggests that conflating causal structure with signalling might be unwise.  


Problems arise if the quantum fields act back on the gravitational field via the stress-energy tensor.  It is relatively easy to see that this must bound the minimum space time four-volume that can be used to localise a measurement outcome and justify our claim that spacetime events are measurement outcomes. This is because all physical measurements take some time and occupy some three-volume. There is a lower bound to this. If a measurement takes place too fast, or is to spatially confined, then the Heisenberg uncertainty principle implies a huge fluctuation of the stress-energy tensor. At some point a black hole is created and the measurement event is causally disconnected from every other observer. This has a physical implication for causal set theory. In that approach, the number of distinct measurement event defines a spacetime volume.  If measurement back-action is taken into account, there is a maximum event density in spacetime, and an effective stochastic discreteness to spacetime volumes.

\section{Gravitational decoherence.}
\begin{figure}[h]
   \centering
 \includegraphics[scale=0.5]{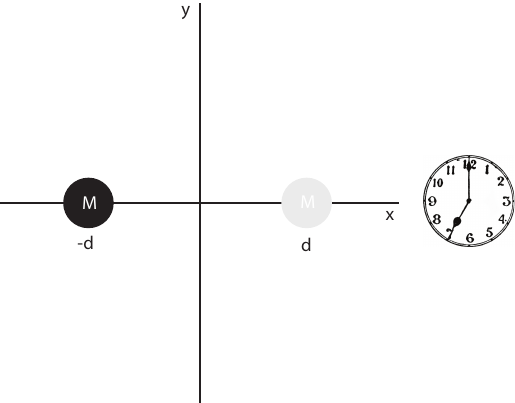} 
 \caption{A single mass is in a superposition of two locations with respect to a fixed coordinate frame (the earth, say).  A very sensitive clock is placed to one side. In a series of repeated trials with exactly the same conditions, the clock will experience two unequal redshifts, randomly choosing one or the other in each trial. As the clock is the only way we can infer the existence of a gravitational field in this setting, we conclude that the the gravitational field at the clock is fluctuating. It is not stationary.     }
 \label{cat-mass-clock}
\end{figure}

The fundamental problem in quantum gravity is this: if a single massive object is prepared in a superposition of two different locations with respect to a fixed coordinate frame, the resulting gravitational field must be non stationary.  An example is shown in Fig. (\ref{cat-mass-clock}).
If instead of the same superposition state for the source mass being used in every trial we placed the mass at one or the other position at random from one trial to the next, the clock will experience the same random red shifts, however in this case it is obvious that the gravitational field is non stationary; we keep changing it from one trial to the next.  In more technical terms, the clock experiment cannot distinguish between an initial pure superposition state, which is a zero entropy state,  and a maximally mixed state with non zero entropy. We might distinguish the two cases using the names  `pure quantum gravity' versus 'stochastic classical gravity'. 

Penrose proposed that we can never actually prepare the pure state required for the first experiment as it will spontaneously collapse into one location or the other as described by the random mixture of the second experiment.  He did not give an explanation for how this can happen.   Many people have now devised experiments\cite{tabletop-qg} that could in principle distinguish the pure state from the classical mixture. These experiments could distinguish pure quantum gravity from stochastic classical gravity. 

The reason is simple:  pure quantum gravity can become entangled --- quantum correlated --- with internal degrees of freedom of particles whereas stochastic classical gravity cannot, it can only be classically correlated. Conversely, stochastic classical gravity can  control quantum systems but cannot entangle them. An entanglement witness can reveal the difference. Any attempt to erase which path information in an entanglement witness, without acting on gravitational degrees of freedom, will fail.  

\begin{figure}[h!]
	\centering
	\includegraphics[scale=0.5]{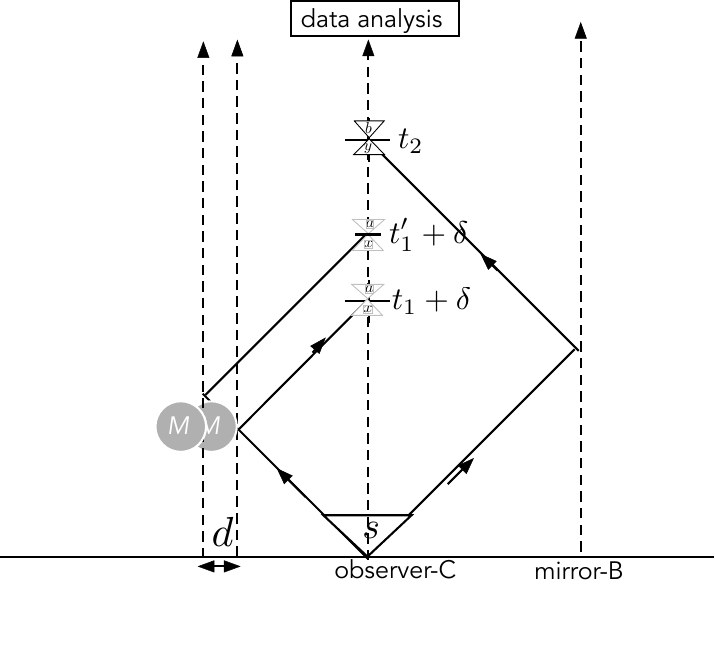}
 \caption{A scheme for using time-bin entanglement and a single agent Bell test to search for gravitational decoherence. }
 \label{q-gravity-one-observer}
	\end{figure}

\begin{figure}[h]
\centering
\includegraphics[scale=0.75]{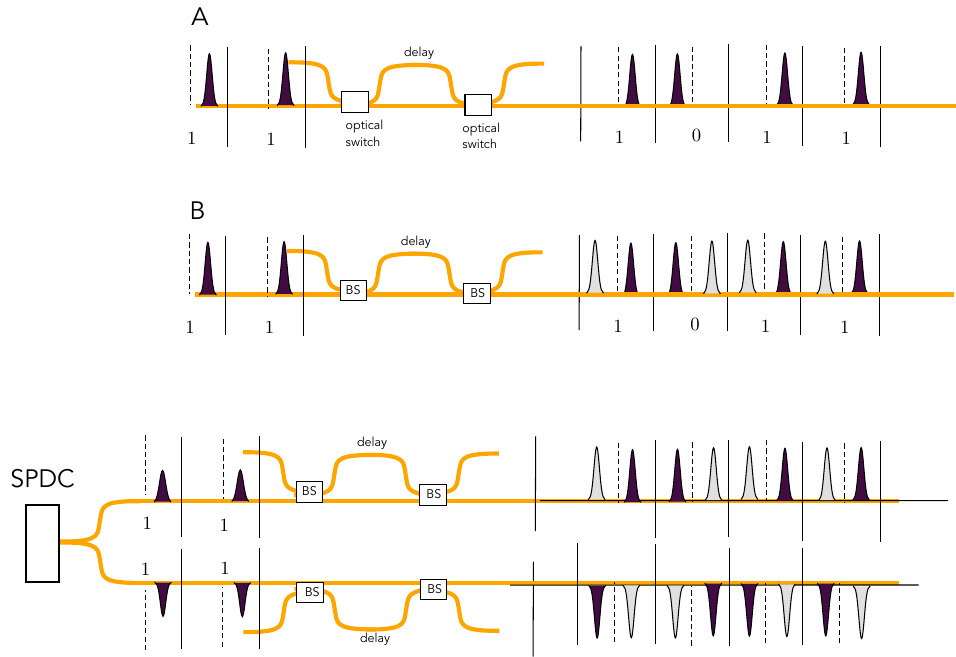}
\caption{[A] A classical optical circuit to encode a bit string into pulse time code. An incoming stream of equally spaced coherent pulses is optically switched onto a direct path or a delay path to encode bits as early or late pulses in each time bin. [B] A time-bin qubit encoder. A sequence of single photon pulses is  input to  an optical circuit using 50/50 fibre couplers to create an equal quantum superposition of logical bits. This is a time-bin encoded qubit.  Note that the total photon number in each time bin is one. [C] A single down conversion source creates pairs of photons simultaneously. Each path passes through a single qubit gate which creates pairs of photons in which one is delayed with respect to the other but we do not know which.   }
\label{time-bins-entanglement}
\end{figure}

Let us return to the question of the gravitational field of a non classical source, such as a large mass in a superposition of two displacements with respect to the rest frame of a source of Bell pairs, see Fig. (\ref{q-gravity-one-observer}).  
 The photon travelling to the left will now return to be detected at two possible times. The Shapiro shift is the same as the mass is the same. If the photon samples a fluctuating gravitational field, the Shapiro shift will be stochastic from pulse to pulse.  
 In order to consider the Shapiro effect on a Bell test we switch to time-bin entangled photons\cite{time-bin-Gisin} rather than polarisation. We can use a version of time dependent multiplexing to encode a qubit into a sequence of single photon pulses, see Fig. (\ref{time-bins-entanglement}).
There will be no change to a Bell violation, provided the temporal and gravitational degrees of freedom do not become entangled.  A time-bin entanglement Bell test could then be an entanglement witness. If the temporal history of the photons does not become entangled with the gravitational field, the Bell violation can be maximal, and otherwise it is reduced. Note that in this single agent bell test, the operations required by the agent are entirely local.

\section{Ringworld: a toy model}
A common theme in the preceding discussion is that a Bell test can be violated by a single agent with a local clock and local interventions (preparation/measurement).  The global structure of spacetime is unknown to the agent; it only has a view from the inside.  When thinking about such experiments, it is hard to  take the inside view of the agent and not import our intuitions of the view from the outside. We will now present an 'intuition pump' which will help you move towards a view `from the inside out.'

\begin{figure}[h!]
	\centering
	\includegraphics[scale=0.5]{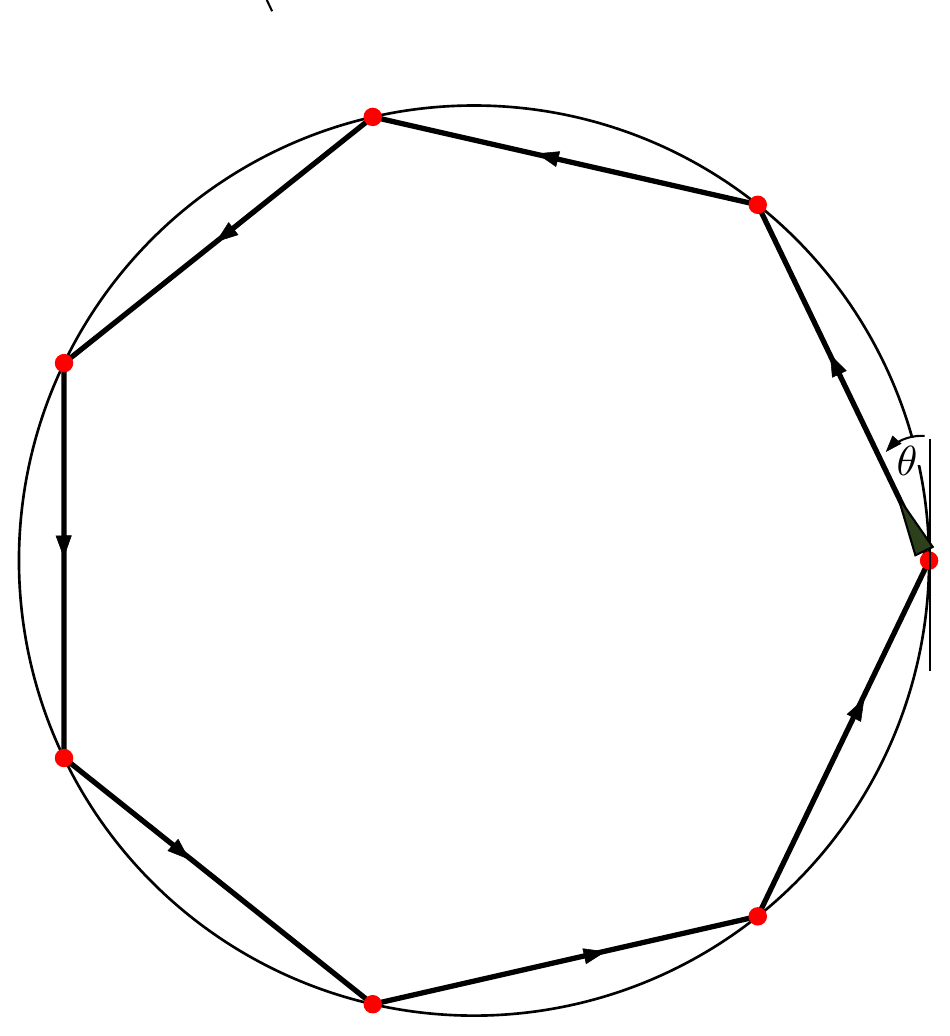}
    \caption{ A agent on the edge can emit light pulses in different directions. They are reflected from the edge and return to the agent. The agent can control the direction using internal actuators.}
 \label{new-ring-world-flat}
	\end{figure}

Suppose a learning agent is moving on a unit circle with a reflecting boundary.  Inside each agent is a clock and a gyroscope. The agent emits a light pulse at each tick of internal clock. This determines the rate $r$ of pulse emission. The gyroscope estimates the direction of its pointer as a function of the ticks of the internal clock. An example of possible configurations is shown in Fig.(\ref{new-ring-world-flat}).  Thus $\theta=\pi/2$ is along a radius and orthogonal to the tangent to the circle at the agent.  We will assume that $\theta=k\frac{\pi}{2K}$, for some integers $k=0,1,\ldots K >>1$. We will set the speed of light, $c=1$ and use units of length such that the radius of the circle is unity. 

The gyroscope is a simple sensor that measures angular accelerations of the agent’s head direction. When combined with a clock, it enables the agent to keep a record of head direction.  The agent can only `know’ its internal states indexed by ticks of the internal clock.

\begin{figure}
    \centering
    \includegraphics[width=0.5\linewidth]{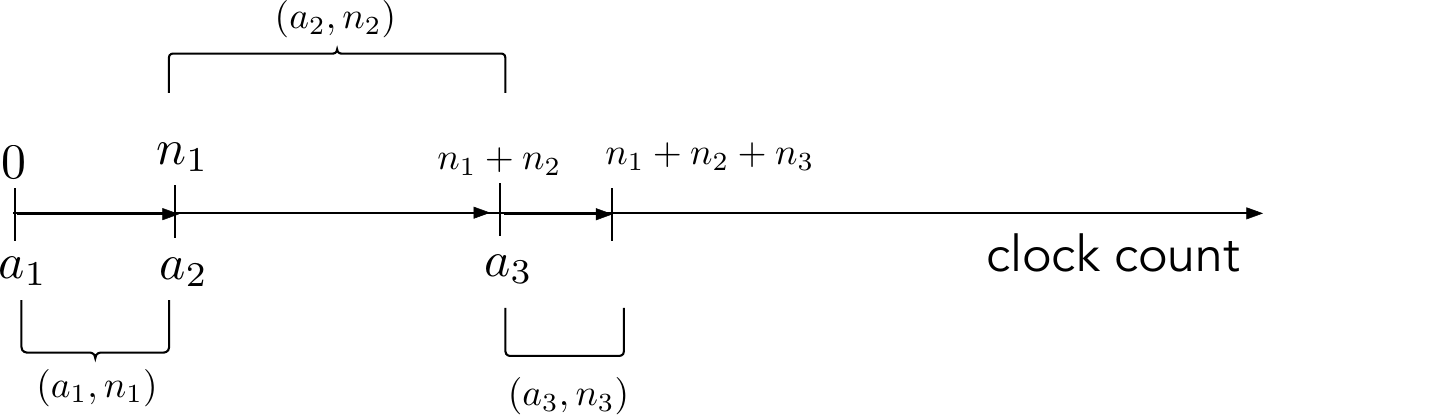}
    \caption{The scheme of agent interventions (head angle at emission ) and sensations (clock count at pulse reception) versus clock count together with the data training pairs sent to the internal learning machine}
    \label{agent-pulse}
\end{figure}

We will assume that the agent emits pulse immediately after receiving a pulse: if a pulse is received at a clock count of $n$, a pulse is emitted at the same clock count $n$.  The head direction at each step determines the clock count of the {\em next} emission.  Each emission event corresponds to a distinct internal state labeled by a setting of its internal gyroscope,  $a$, and a reading of the clock at pulse emission, $n$. Our protocol means that a pulse received at clock count of  $n$  is a pulse returning from the previous emission when the head direction was $a$.   The only thing the agent has access to are these two things. An internal state is an ordered pair of numbers  $S=(a, n)$ where  $n$ is the clock count for the next emission, and  $a$ is the record of head direction at the previous emission.   This is summarised in Fig. (\ref{agent-pulse})

We now equip the agent with an internal learning machine. It works like this. At each tick of the internal clock, the learning machine is sent the head direction, $a$.  It then quickly tries to predict when the next light pulse will be received by generating an integer $N$. For simplicity, we will assume there are only four settings for $a$. These are labeled with angles given by ($0.01\pi, 0.1\pi,0.15\pi, 0.2\pi$, but the agent does not know this. It only knows that there are four different headings, as recorded by its internal gyroscope.  At each step one of these four headings is chosen at random. 

\begin{figure}
    \centering
    \includegraphics[width=0.5\linewidth]{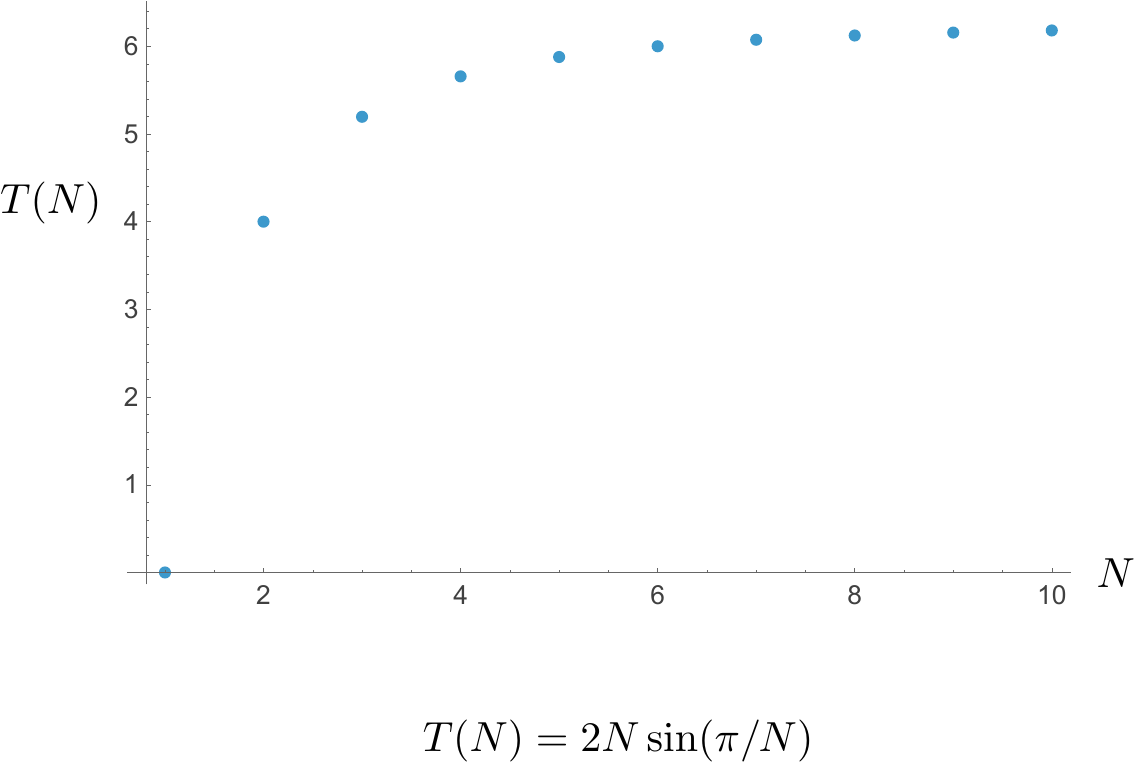}
    \caption{A plot of how return time of a pulse varies with head angle. Arbitrary units for time. }
    \label{new-path-length}
\end{figure}
\begin{figure}
    \centering
    \includegraphics[width=0.65\linewidth]{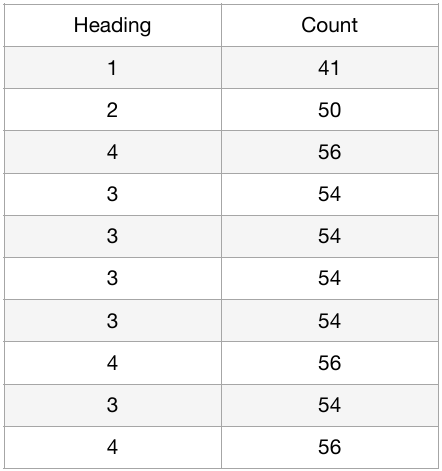}
    \caption{A sample of typical data, actions (head angle) and sensations (count to received pulse),  used as inputs to the learning machine. Head directions are settings of an internal gyroscope and chosen at random. }
    \label{data}
\end{figure}

Assume that the agent is stationary. Given our external god-like view of the agent’s world, from the outside in, we can easily give a relation between head angle and time taken for a pulse to return to the agent. We assume that spacetime is flat inside the disk. The time taken is then given by $T(\theta)=40(\cos\theta+\sin\theta)$.
We have used an arbitrary scale for time units. The shortest period occurs for a pulse sent along a diagonal and the longest for a pulse sent at 45 degrees to the diagonal. See Fig. (\ref{new-path-length}).
The agent does not know this function. It simply generates a list of ordered pairs, a label for the head angle and number of ticks of the clock until the pulse returns. The data can be displayed like the table in Fig. (\ref{data}). 

\begin{figure}
    \centering
    \includegraphics[width=0.5\linewidth]{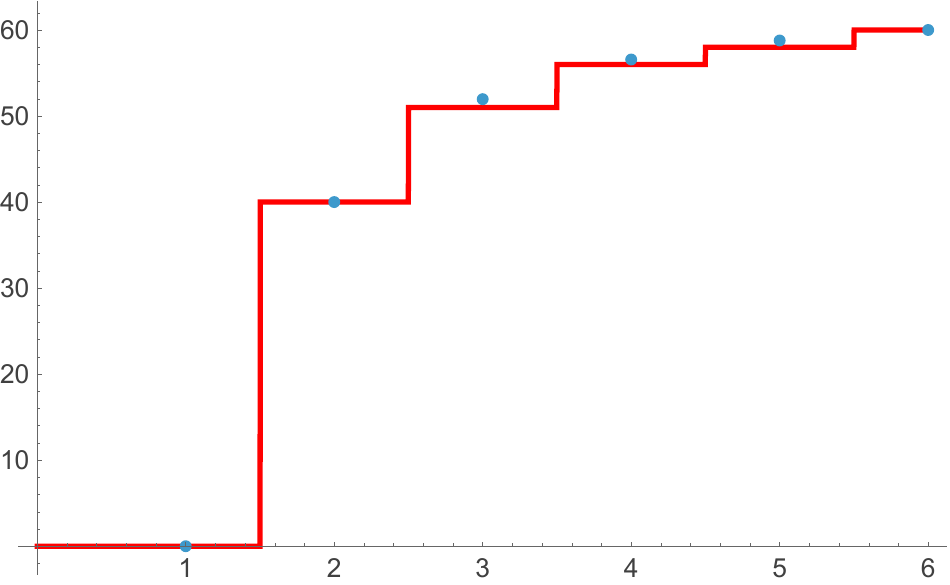}
    \caption{An example of the prediction made by a learning machine inside the agent using a long data set of the kind show in Fig (\ref{data}). Using only four randomly chosen head directions. The blue dots represnet the ground truth function that agent is trying to learn.}
    \label{nn-prediction-flat}
\end{figure}

The physical machine that implements the learning could be specified in various ways. This can be almost anything, for example a physical neural network or a physical restricted Boltzmann machine. We will assume for illustration that it is machine that implements a neural network algorithm and that it has a very large number of examples to train on. In Fig.(\ref{nn-prediction-flat})  We plot an example of the predictions made by this machine once it has been trained using 1000 training pairs.  
The learned function is stored as physical settings of the physical learning machine inside the agent. In so far as it has learned this function it has learned a proxy for Euclidean geometry. 

The case of head angles that are irrational multiples of $\pi$ requires special consideration. Orbits do not close, but almost-returns are certain. The time between them is on average 
$1/\epsilon$ steps for phase space arc length $\epsilon$ ...  `width' of agent. For most irrational $\theta/\pi$, the sequence of recurrence times to $epsilon$-arcs behaves as:
$$
Pr(T>t)\sim e^{-\epsilon t}
$$
 where $t$ is the number of bounces. 

In this world, the agent is the only source of light. If there are extrinsic sources of light pulses that do not originate from agents, the training data of the agent is corrupted. Occasionally one of these pulses will be received by the agent. How does the agent distinguish these `background’  pulses from those that the agent itself emits?

From the agent’s point of view, these random emissions are `background noise’.  A key feature of learning is the ability to cope with some uncertainty in the data. In this case, the sensor records are not perfectly correlated with the recordings of head direction. How much noise can be tolerated before learning begins to degrade?

\begin{figure}
    \centering
    \includegraphics[width=0.5\linewidth]{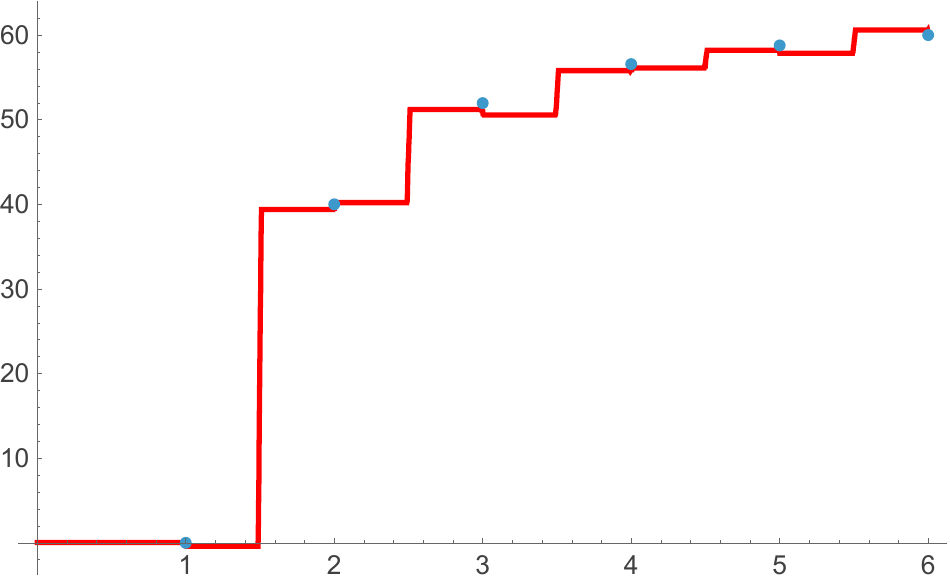}
    \caption{A comparison of the unknown function and the learned function when extrinsic light pulses corrupt the training data. The number of training samples is 1000. }
    \label{nn-prediction-flat-error}
\end{figure}

If it receives a random light pulse, the correlation inherent in the ordered pairs  $(a_j,n_j)$ is contaminated by independent errors in the  $n_j$.   We can represent this by adding/subtracting a small random integer,$e_j$ , to each count component
\begin{equation}
    (a_j,n_j)\rightarrow (a_j,n_j+e_j),\ \ \ |e_j|<n_j
\end{equation}
The emission rate from background sources compared to the agent’s emission rate is an important fact. If it is small, we might expect the agent can still learn how to control the return time of pulses with changing head direction. Assume $e_j$  is a random integer between -4 and 4. In Fig. (\ref{nn-prediction-flat-error}),  1,000 trials are used to lean the unknown function. The prediction remains quite good. 

\begin{figure}[h!]
    \centering
    \includegraphics[width=0.75\linewidth]{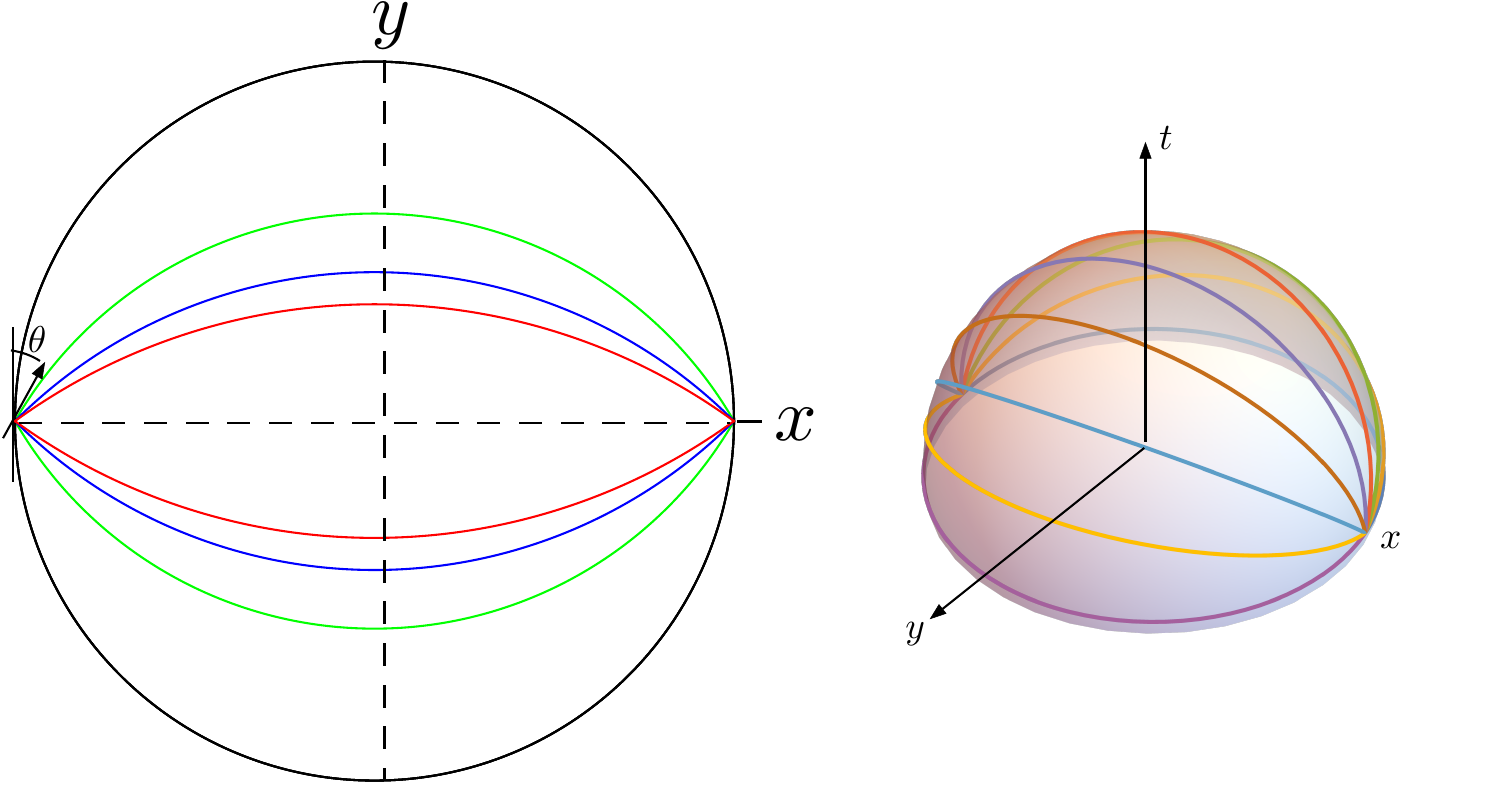}
    \caption{The interior of the disk is composed of a spatially varying refractive index that decreases away from the centre. Red: $\theta=\pi/5$, Blue: $\theta=\pi/4$, Red=$\theta=\pi/3$}
    \label{fish-eye2}
\end{figure}

We now turn to a non Euclidean example. The analog of de Sitter space is the Poincar\'{e} disk, while the anti-de Sitter space is equivalent to Maxwell’s fish-eye lens\cite{PhysRevA.102.023528}. We treat the latter here.  The  Maxwell fisheye lens, in the  disk of radius $R$, has a radially dependent refractive index, 
\begin{equation}
	n(r)=\frac{2}{1+(r/R)^2}
\end{equation}
We will assume that $R=1/2$ and thus the refractive index is equal to one on the boundary. Light rays propagating within an infinite two-dimensional plane lens, rays trace out perfect circles. In the case of a disk, rays starting on the boundary are focused at the antipodal boundary point, see Fig (\ref{fish-eye2}). Each curve is defined by a head angle as in the flat space case. 
The situation for a reflecting circular boundary, of radius $1/2$ centred at the origin, is treated in \cite{Lukin}. 

In \cite{PhysRevA.102.023528} it is shown that the null geodesics on the sphere of radius $R$ correspond to the light rays in the Maxwell fisheye disk, see Fig.(\ref{fish-eye2})   
This implies that the time taken for a light pulse to travel from one side of the disk to the other is independent of the head direction, unlike the flat space case previously discussed.

Each agent is equipped with an internal `gyroscope' that determines the head direction $\theta$ and an internal `clock' that counts the time taken for a light pulse to be returned.  In the case of the empty disk, the time taken depends on the head direction, and the agent can learn the relationship given a simple internal learning machine. In the case of the Maxwell fish-eye disk, the time taken is independent of head direction. These are the two `laws of physics' that the agents learn for each case.  In both cases the law can be learned using only local actuators and sensors inside the agent. It knows nothing abut the propagation of light from an external, god-like, view.  It projects the learned law `from the inside out'.

\section{Conclusion.}
Einstein constructed general relativity using a profound intuition about how we use local clocks and rulers, yet was a little vague about just how such things were built from the fundamental theory. In a quantum world, local clocks and rulers are replaced with local measurements made with quantum sensors. The events from which spacetime is constructed must be classical measurement results made on quantum fields. 

If gravity is spacetime, then we must regard gravity as sharing the features of quantum measurements: irreducible stochasticity and contextuality.  However there is a catch here: it is very difficult to conceive of measurements without positing a background spacetime. In the usual test of Bell inequalities with three agents, two of whom are space-like separated, we simply assume that the measurement devices 'have' spacetime coordinates. Yet the observed violation of the Bell inequality is independent  of the spacetime coordinates of the detectors involved. We described a single-agent Bell experiment the results of which are the same as the usual three-agent scenario. When the agents are space-like separated we feel uneasy, but we should feel just as uneasy when they are time-like separated as that case seems to imply local retro-causality.  

We have argued that in thinking about how to make gravity consistent with quantum theory, we should take an inside-out view of the world, an agent centric view in which notions of global time and space are secondary, if not entirely illusory.

\section*{Acknowledgments.}
I wish to thank Peter Evans for useful discussions.

\bibliography{spacetime}

\begin{thebibliography}{10}

\bibitem{Buzaski}
G.~Buzs\'{a}ki.
\newblock {\em The Brain from Inside Out}.
\newblock Oxford University Press, 2019.

\bibitem{Dewitt2011TheRO}
C{\'e}cile~Morette Dewitt and Dean Rickles.
\newblock The role of gravitation in physics.
\newblock In {\em Report from the 1957 Chapel Hill Conference}, 2011.

\bibitem{Gisin2013}
Nicolas Gisin.
\newblock {\em Are There Quantum Effects Coming from Outside Space--Time? Nonlocality, Free Will and ``No Many-Worlds''}, pages 23--39.
\newblock Springer New York, New York, NY, 2013.

\bibitem{PhysRevD.96.105004}
T.~G. Downes, J.~R. van Meter, E.~Knill, G.~J. Milburn, and C.~M. Caves.
\newblock Quantum estimation of parameters of classical spacetimes.
\newblock {\em Phys. Rev. D}, 96:105004, Nov 2017.

\bibitem{Scarani2019}
Valerio Scarani.
\newblock {\em Bell Nonlocality}.
\newblock Oxford Graduate Texts. Oxford University Press, 2019.

\bibitem{shapiro}
Irwin~I. Shapiro.
\newblock Fourth test of general relativity.
\newblock {\em Phys. Rev. Lett.}, 13:789--791, Dec 1964.

\bibitem{Einstein1970}
Albert Einstein.
\newblock Autobiographical notes.
\newblock In Paul~A. Schilpp, editor, {\em Albert Einstein: Philosopher-Scientist}, pages 1 -- 94. Harper \& Row Publishers, New York, 1970.

\bibitem{Kempf2021}
A~Kempf.
\newblock Replacing the notion of spacetime distance by the notion of correlation.
\newblock {\em Front. Phys}, 2021.

\bibitem{causalset}
S.~Surya.
\newblock The causal set approach to quantum gravity.
\newblock {\em Living Reviews in Relativity}, 22, 2019.

\bibitem{tabletop-qg}
E.~Adlam.
\newblock Tabletop experiments for quantum gravity are also tests of the interpretation of quantum mechanics.
\newblock {\em Found Phys}, 52:115, 2022.

\bibitem{time-bin-Gisin}
Mattaeus Halder, Alexios Beveratos, Nicolas Gisin, Valerio Scarani, Christoph Simon, and Hugo Zbinden.
\newblock Entangling independent photons by time•measurement.
\newblock {\em Nature Physics}, 3, 2007.

\bibitem{PhysRevA.102.023528}
Huanyang Chen, Sicen Tao, Jakub B\ifmmode~\check{e}\else \v{e}\fi{}l\'{\i}n, Johannes Courtial, and Rong-Xin Miao.
\newblock Transformation cosmology.
\newblock {\em Phys. Rev. A}, 102:023528, Aug 2020.

\bibitem{Lukin}
J.~Perczel, P.~K\'om\'ar, and M.~D. Lukin.
\newblock Quantum optics in maxwell's fish eye lens with single atoms and photons.
\newblock {\em Phys. Rev. A}, 98:033803, Sep 2018.

\end{thebibliography}

\end{document}